# Profound impacts of interlayer interactions in bilayer altermagnetic $V_2S_2O$

Siqi Xu[1], Qilong Cui[1], Shaowen Xu[2], Xianbo Chenwei[1], Jiahao Zhang[1], Ruixue Li[1], Yuan Li[1], Gaofeng Xu[1†], and Fanhao Jia[1*]

[1]Department of Physics, Hangzhou Dianzi University, Hangzhou 310018, China

[2]School of Physics and Optoelectronic Engineering, Hangzhou Institute for Advanced Study, University of Chinese Academy of Sciences, Hangzhou 310024, China

*Contact author: fanhaojia@hdu.edu.cn
†Contact author: xug@hdu.edu.cn

## ABSTRACT

Two-dimensional altermagnets exhibit exceptional potential for low-power spintronics via nonrelativistic spin splitting and zero net magnetization. Here, we systematically investigate the influence of interlayer interactions on the electronic, magnetic and quantum transport properties of bilayer vanadium oxysulfide ($V_2S_2O$) — a prototypical layered altermagnet — using density functional theory and non-equilibrium Green's function calculations. Our results reveal that interlayer interactions predominantly modulate the p-orbital derived top valence bands, inducing a profound competitive valence band maximum position between $\Gamma$-point $p_z$ and X/Y-point $p_{xy}$ orbitals, with an energy difference as small as 9 meV. Furthermore, interlayer interactions suppress the piezomagnetic effect and impose additional requirements on the type of strain for the bilayer system, compared to its monolayer counterpart. Out-of-plane external electric fields effectively weaken interlayer coupling by enlarging the energy difference of $\Gamma$/X-Y top valence bands to 170 meV. Quantum transport simulations on a bilayer Au/$V_2S_2O$/Au two-probe device demonstrate the presence of pronounced spin current. Interlayer interactions reduce the transmission spin polarization from nearly 100% (monolayer) to ~60% (bilayer) for energies above the Fermi level. Notably, gate-voltage modulation exhibits significant asymmetry in controlling charge-to-spin current conversion efficiency, originating from the out-of-plane symmetry breaking induced by the electrode geometry. Specifically, a positive gate voltage markedly enhances the contribution of the bottom layer to the overall spin polarization, while a negative gate voltage induces a marginal reduction of transmission spin polarization, attributed to the inherently weak polarization contribution of the bottom layer. These findings provide essential insights for the design and optimization of multilayer altermagnetic spintronics.

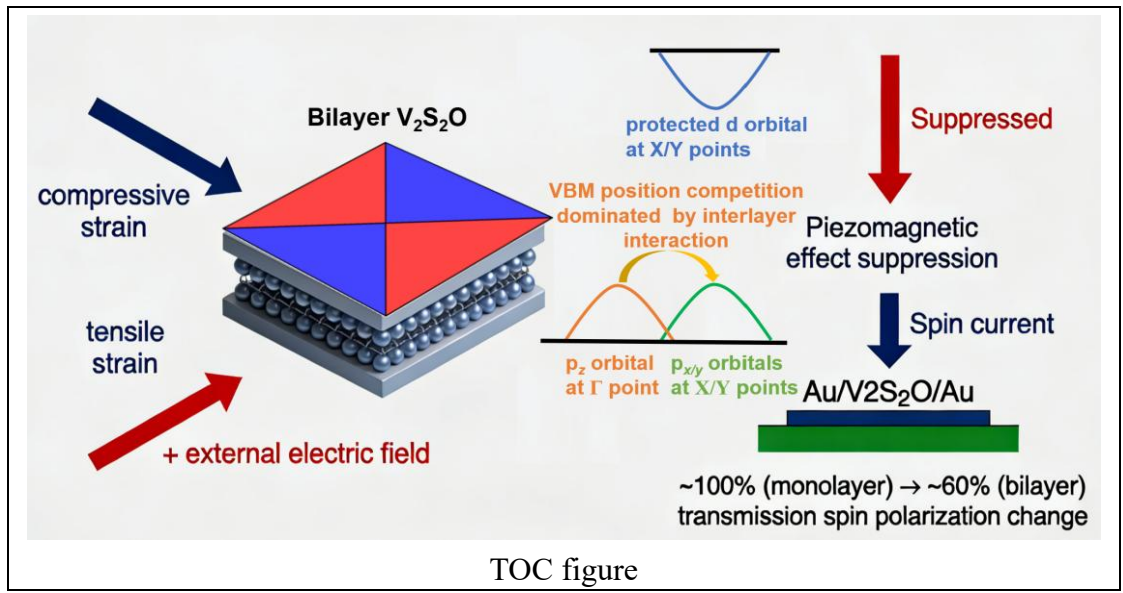
TOC figure

The long-standing pursuit of low-power, highly integrated spintronic devices has driven the search for magnetic materials that transcend the limitations of conventional ferromagnets and antiferromagnets [1-3]. Ferromagnets enable spin polarization but are constrained by inherent stray fields that limit miniaturization and integration. Antiferromagnets, by contrast, allow for miniaturization and fast, low-power operation; however, the intrinsic spin splitting essential for efficient spin transport is absent [4]. Altermagnetism, an emerging collinear magnetic order characterized by momentum-dependent nonrelativistic spin splitting and zero macroscopic magnetization, has emerged as a game-changing solution to this dilemma [5-7]. By breaking combined Parity-Time symmetry without net magnetic moment, altermagnets combine the functional advantages of ferromagnets (spin-polarized transport) with the stability of antiferromagnets (no stray fields) [8], making them ideal candidates for next-generation spintronics [9-11].

Layered two-dimensional (2D) altermagnets have attracted particular attention due to their exceptional tunability and compatibility with nanoscale device architectures [12-13]. Classic theoretical studies established that altermagnetic spin splitting arises from symmetry breaking and crystal rotational symmetry coupling between spin sublattices, rather than spin-orbit coupling or net magnetization [7]. Experimental breakthroughs, such as the observation of giant spin splitting in CrSb thin films [14] and anomalous Hall effect in $RuO_2$ [15], have validated the potential of altermagnets in spintronic applications. Recent advances further demonstrate that layered altermagnets can achieve remarkable device performance: for instance, $KV_2Se_2O$-based [16-17] altermagnetic tunnel junctions exhibit tunnel magnetoresistance ratio exceeding $10^3$% [18], far surpassing traditional ferromagnetic tunnel junctions [19-20]. Additionally, altermagnet-topological superconductor heterostructures have been theoretically predicted to support quantized crossed Andreev reflection [21], opening avenues for topological quantum computing in zero magnetic fields [22].

Despite these progresses, critical gaps remain in understanding layered altermagnets, especially in multilayer systems where interlayer interactions introduce new degrees of freedom [23-24]. Monolayer altermagnets, such as $V_2S_2O$, have been shown to exhibit giant spin splitting and spin-valley locking, with spin transport properties dominated by orbital-selective contributions from p and d orbitals [13]. However, transition from monolayers to bilayers—an essential step for device fabrication—introduces interlayer coupling effects that can drastically reshape electronic structure and transport behavior [25]. While external stimuli such as strain [26] and electric fields [27] serve as effective tools for tailoring 2D materials—e.g., breaking valley degeneracy and enhancing piezomagnetic effects [28-29] in monolayers via strain, or inducing Stark shifts and symmetry breaking via electric fields—their impact in bilayer systems remains less clear due to the specific role of interlayer coupling. Previous studies on related layered magnets suggest that interlayer interactions can modulate band alignment , crystal symmetry [30], and magnetic order [31]. Yet, their specific influence on altermagnetic spin transport—particularly regarding orbital competition and responsiveness to external fields—remains largely unexplored. Furthermore, device geometry, such as the use of top-capped electrodes, may introduce asymmetric layer contributions to spin transport, leading to directional dependence in gate-voltage modulation. This effect also requires systematic investigation.

Here, we systematically investigate the electronic, magnetic, and quantum transport properties of bilayer altermagnetic $V_2S_2O$ using density functional theory (DFT) and non-equilibrium Green's function (NEGF) calculations. We focus on the profound impacts of interlayer interactions, exploring orbital competition, strain/electric field tunability, and asymmetric spin transport. Our results reveal that interlayer coupling induces a delicate valence band maximum (VBM) competition between Γ-point $p_z$ and X/Y-point $p_{xy}$ orbitals, modulates external field responses, and significantly reduces transmission spin polarization. We further demonstrate that vertical gate voltage exhibits asymmetric modulation of charge-to-spin conversion efficiency. These findings provide fundamental insights for designing high-performanc layered altermagnets, externally tunable spintronic devices.

**RESULTS AND DISCUSSIONS**

**Interlayer Coupling-Induced Orbital Competition**
Layered $V_2S_2O$ adopts cubic lattice with a point group symmetry of $D_{4h}$. Its monolayer comprises three atomic layers, with fully flat V–O plane sandwiched by two outer S layers (Fig. 1(a)). The magnetic $V^{3+}$ adopts octahedron coordination with the O and S atoms, where the significantly shorter axial V–O bonds (1.92 Å) and the stronger crystal-field strength of O compared to S (V–S bonds 2.47 Å) result in a tetragonally compressed octahedron (axial compression). This structural distortion lifts the degeneracy of the ideal octahedral ($O_h$) orbital splitting ($t_{2g}$ and $e_g$), destabilizing states with a finite z-component, as shown in Fig. 1(b). According to Hund's rules, the two electrons of $V^{3+}$ occupy the two lowest-energy orbitals $d_{xy}$ and $d_{x^2-y^2}$ with parallel spins, giving a spin-triplet ground state (S = 1). In the magnetic ground of monolayer $V_2S_2O$, the nearest-neighbor V ions adopt antiferromagnetic (AFM) order. Due to the specific crystal symmetry ($D_{4h}$) and the orbital-anisotropic crystal-field ground state (occupied $b_{2g}$ and $b_{1g}$ orbitals), the spin-splitting of electronic bands is not simply forbidden as in conventional centrosymmetric antiferromagnets. Instead, the alternating spin polarization becomes momentum-dependent, leading to spin-split bands in momentum space where states of opposite spin are connected by crystal-symmetry operations (e.g., a combination of translation and rotation). This results in a non-relativistic, symmetry-enforced spin splitting in the absence of net magnetization—a hallmark of altermagnetism.

Figure 1(c) shows the spin resolved band structure of monolayer $V_2S_2O$, which presents a giant spin splitting. This splitting is highly anisotropic and momentum-dependent, with a wide range of dominant effects, from -0.45 eV to 1.55 eV. The bands along Γ-M are spin degenerate due to the spin group symmetry of $|C_2||C_{4z}|$, whereas the bands along Γ-X and Γ-Y are identical in energy but present spin splitting. The X and Y points in this system are donated as the band valleys, where both VBM and conduction band maximum (CBM) are located. For each valley, they belong to opposite spin channel, driven by the distinct local chemical environments of the respective sites.

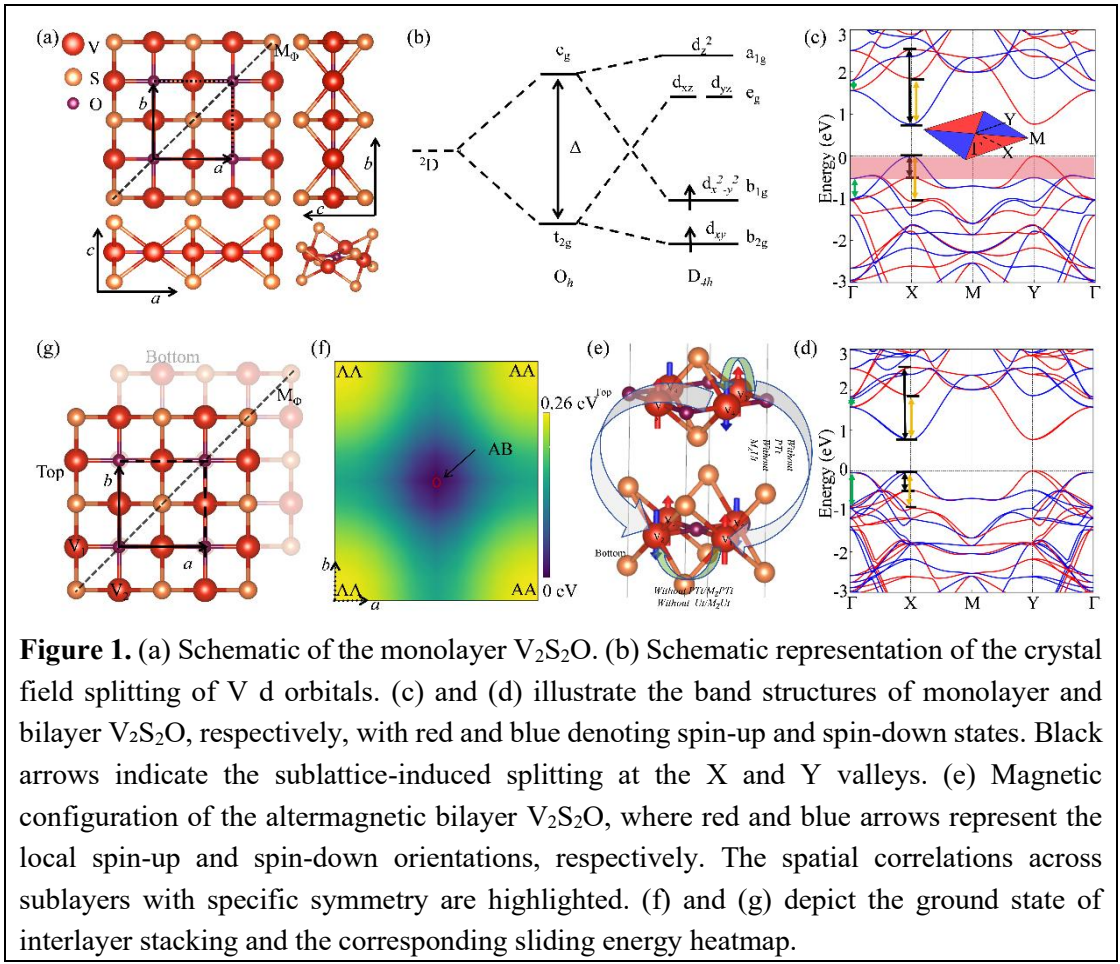

**Figure 1.** (a) Schematic of the monolayer $V_2S_2O$. (b) Schematic representation of the crystal field splitting of V d orbitals. (c) and (d) illustrate the band structures of monolayer and bilayer $V_2S_2O$, respectively, with red and blue denoting spin-up and spin-down states. Black arrows indicate the sublattice-induced splitting at the X and Y valleys. (e) Magnetic configuration of the altermagnetic bilayer $V_2S_2O$, where red and blue arrows represent the local spin-up and spin-down orientations, respectively. The spatial correlations across sublayers with specific symmetry are highlighted. (f) and (g) depict the ground state of interlayer stacking and the corresponding sliding energy heatmap.

The spin splitting at CBM and VBM have different natures. We use green, yellow and black double-arrows to indicate the orbital splitting at Γ point, exchange-splitting and sublattice-splitting at X point (the same at Y point). The spin splitting at CBM (~1.1 eV) is derived from the exchange splitting, while the spin splitting at VBM (~0.5 eV) is derived from the spin sublattice splitting. This intrinsic difference makes CBM and VBM present distinct spin splitting response with respect to external strain and electric field. Moreover, the bottom conduction bands are mainly derived from $d_{xz}$ and $d_{yz}$ orbitals, while the top valence bands are mainly derived from p orbitals. The top valence bands at Γ point are mainly derived from $p_z$ orbital, while the VBM at X or Y points are derived from $p_x$ and $p_y$ orbitals, respectively. Thus, the top valence bands at Γ with z-component orbital are expected to be more significantly affected by interlayer interaction, indicated by the band structure of bilayer $V_2S_2O$ as shown in Fig. 1(d).

Figure 1(e) displays ground-state stack structure of bilayer $V_2S_2O$. It adopts an AB-like stacking [32-33] with a relative lower energy of 0.26 eV/f.u. with respect to the AA stacking (Fig. 1(f)). The magnetic ground state of bilayer adopts a G-type AFM arrangement (Fig. 1(g)). In Fig. 1(d), the impacts of interlayer interactions of bottom conduction bands in a relatively moderate magnitude, due to they are mainly derived from the d orbitals which are protected within the octahedral. On the contrary, interlayer interactions are significant for those more extend p orbitals. For example, at the Γ point, the $p_z$ orbitals of top valence bands splits into an orbital close to the Fermi level (~-0.05 eV) and an orbital 0.9 eV below the Fermi level. In comparison, the VBM at X/Y points that are mainly derived from $p_x$ and $p_y$ orbitals are as robust as the conduction bands against the interlayer interactions, where both sublattice splitting and orbital splitting are maintained. It's worth mentioning that monolayer $V_2S_2O$ has a well-defined direct band gap at X/Y points, while bilayer $V_2S_2O$ has a quasi-direct band gap: (i) the CBM and VBM positions slightly deviated from the X/Y points; (ii) interlayer coupling promotes a profound competition of VBM position between $p_z$ orbital at the Γ point and $p_{xy}$ orbitals at the X/Y points. The relative energy difference of the top valence band at the X/Y points and at the Γ point of AB-stacking bilayer $V_2S_2O$ is only 9 meV. Furthermore, in AA-stacking bilayer, stronger interlayer coupling can cause the VBM position to shift from the X/Y points to the Γ point.

**Strain and Electric Field Tuning the Interlayer Coupling**

The diagonal mirror symmetry ($M_Φ$), which protects the degeneracy between X and Y valleys, can be destroyed via uniaxial strain, laying the basis for the piezomagnetic application in altermagnetic systems [34-35]. Figure 2(a) compares the band structures under no strain, uniaxial compressive (-2%), and tensile (2%) strains (along *a*-axis). While these uniaxial strains only slightly reduce the band gap from pristine 0.79 eV to 0.75 eV, they significantly modify the competition of VBM position between $p_z$ orbital and $p_{xy}$ orbitals. Specifically, compressive strain effectively elevates the energy of top $p_{xy}$ orbitals by reducing the sublattice splitting along the X direction ($Δ_X$ and $Δ_Y$ of Fig. 2(b)-(c)). In contrast, tensile strain increases sublattice splitting. Consequently, under compressive strain, the bilayer $V_2S_2O$ possesses a pronounced direct-like band gap, whereas tensile strain renders the system a fully indirect band gap semiconductor.

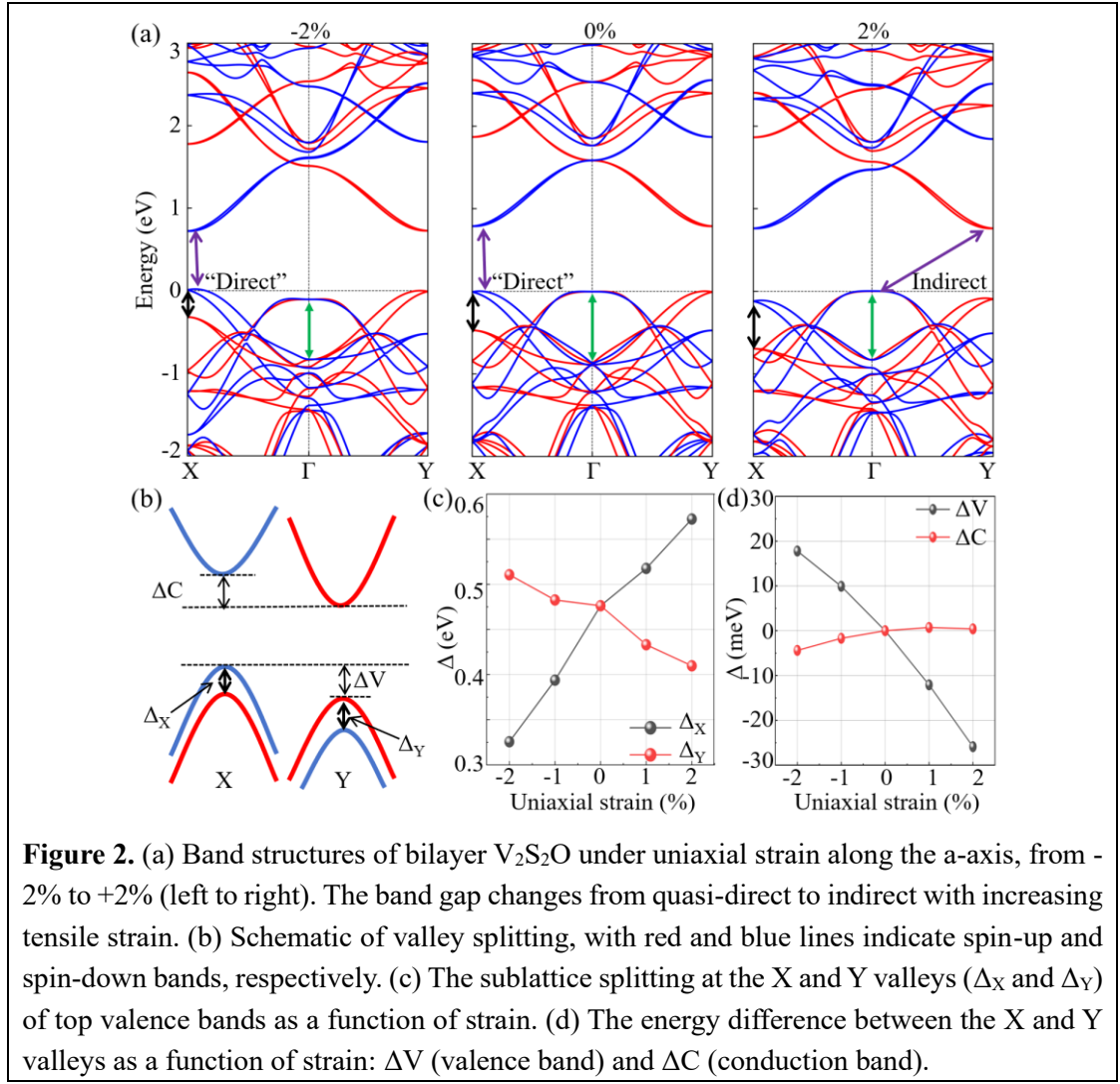

**Figure 2.** (a) Band structures of bilayer $V_2S_2O$ under uniaxial strain along the a-axis, from -2% to +2% (left to right). The band gap changes from quasi-direct to indirect with increasing tensile strain. (b) Schematic of valley splitting, with red and blue lines indicate spin-up and spin-down bands, respectively. (c) The sublattice splitting at the X and Y valleys ($\Delta_X$ and $\Delta_Y$) of top valence bands as a function of strain. (d) The energy difference between the X and Y valleys as a function of strain: $\Delta V$ (valence band) and $\Delta C$ (conduction band).

In addition, the energetic degeneracy lifts derived from uniaxial strain give rise to a pronounced spin–valley polarization (indicated by the $\Delta V$ and $\Delta C$ for valence band and conduction band, respectively). Under strain, the valley polarization of the valence band is much higher than that of the conduction band. As shown in Fig. 2(d), the $\Delta V$ is 18 meV and 27 meV under -2% and 2% strains, respectively. While the values of $\Delta C$ are in the range from the 4 meV of -2% strain to the 1 meV of 2% strain, indicating that the response of $\Delta C$ to strain is weaker. It is worth mentioning that a large valley offset ($\Delta V$ or $\Delta C$) will be beneficial for achieving piezomagnetic effect [26]. Therefore, from the perspective of strain response, hole doping, due to larger $\Delta V$, is more likely to achieve a strong valley piezomagnetic effect in a $V_2S_2O$ system. On the other hand, the tensile strain will change the VBM position from valley points to the $\Gamma$ point, which will greatly suppress piezomagnetic effects. Hence, to achieve a strong piezomagnetic effect in bilayer $V_2S_2O$, we suggest to simultaneously apply both compressive strain and hole doping. This differs from the case of monolayer $V_2S_2O$, where the piezomagnetic effect of hole-doped monolayers does not require consideration of strain type because of the energy window between its valley point and $\Gamma$ point of the top valence band is large enough.

We further investigate the influence of an external electric field (EEF) on bilayer $V_2S_2O$, which offers an alternative way to manipulate its magnetic and electronic properties. Unlike strain, electric field acts directly on the local potential, charge density, potentially leading to distinct switching behaviors. The EEF is perpendicularly applied on the $V_2S_2O$ sheet with an intensity of ±0.1 V/Å. As shown in Figure 3, the EEF lifts the energy degeneracy between layers. The interlayer band gap effectively decreases in proportion to the EEF intensity. The band gap of bilayer $V_2S_2O$ was reduced to 0.28 eV under an EEF of 0.1 V/Å. The d orbitals of conduction bands are extremely sensitive to EEF which directly interacts with the out-of-plane charge distribution (Fig. S2). This results in a large Stark shift, manifested as an overall band shift rather than significant spin splitting changes. The Stark shift of the valley of the conduction band approximately equals to that of the valence band, and can reach up to 530 meV at an EEF of 0.1 V/Å.

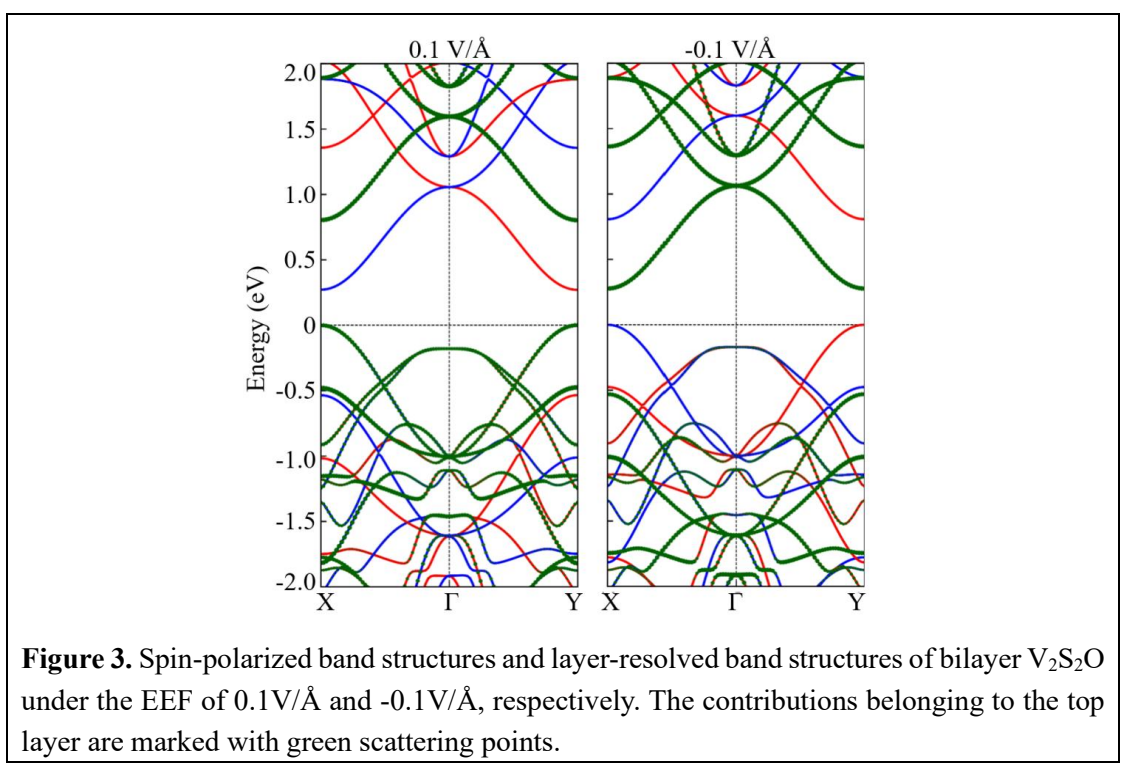

**Figure 3.** Spin-polarized band structures and layer-resolved band structures of bilayer $V_2S_2O$ under the EEF of 0.1V/Å and -0.1V/Å, respectively. The contributions belonging to the top layer are marked with green scattering points.

On the other hand, a negative EEF lifts the overall electronic potential energy of top layer and relatively lowers that of the bottom layer. This causes a relative shift in energy levels of different layers, which were originally degenerate, disrupting the original band alignment. In other words, EEF provide a layer-resolvable way to study the bilayer system. Moreover, EEF can effectively change the competition of the VBM position between its valley points and Γ point, for which the energy window was enlarged from the 9 meV of pristine bilayer to the 170 meV under EEF of 0.1 V/Å. This will greatly enhance the dominant role of the valley electronic state of bilayer $V_2S_2O$ in applications. In a sense, we can claim that EEF can effectively weaken interlayer interactions, allowing the physical properties of the bilayer to return to a state close to that of monolayer.

It's also worth mentioning that applying out-of-plane EEF can turn the normal AFM bilayer $V_2S_2O$ into an altermagnet-like state. The normal AFM bilayer $V_2S_2O$ adopts a C-type AFM configuration (Fig. S3), which preserves the inversion symmetry, thus showing spin degeneracy in moment space. Out-of-plane EEF breaks the inversion symmetry and induces Stark shift between the top layer and bottom layer. In this case, the AFM bilayer $V_2S_2O$ changes to two layer-resolvable altermagnetic layers as well (Fig. S4). The Stark shift caused by EEF, is almost identical in both AFM and altermagnetic cases. However, unlike the pristine altermagnetic bilayer, the AFM bilayer under an EEF also exhibits momentum-dependent spin splitting, but the splitting is reversed between the top and bottom layers.

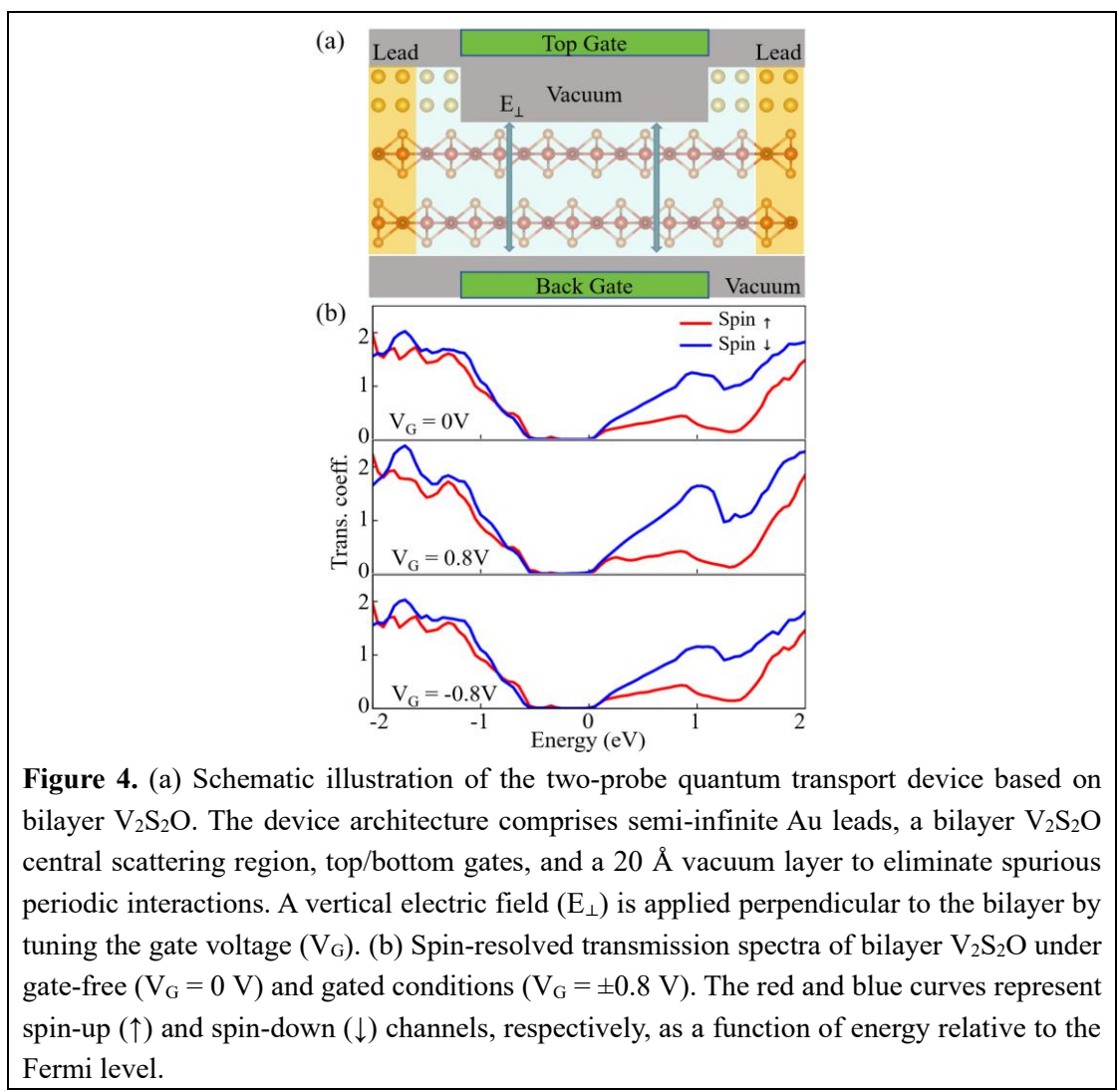

**Figure 4.** (a) Schematic illustration of the two-probe quantum transport device based on bilayer $V_2S_2O$. The device architecture comprises semi-infinite Au leads, a bilayer $V_2S_2O$ central scattering region, top/bottom gates, and a 20 Å vacuum layer to eliminate spurious periodic interactions. A vertical electric field ($E_\perp$) is applied perpendicular to the bilayer by tuning the gate voltage ($V_G$). (b) Spin-resolved transmission spectra of bilayer $V_2S_2O$ under gate-free ($V_G = 0$ V) and gated conditions ($V_G = \pm0.8$ V). The red and blue curves represent spin-up (↑) and spin-down (↓) channels, respectively, as a function of energy relative to the Fermi level.

**Gate-Voltage Modulated Spin Transport**

Due to the pronounced spin-valley locking properties of altermagnetic $V_2S_2O$ in momentum space, it is expected that the transport efficiency of quasiparticles with different spins will differ in specific directions (e.g., along the *x* and *-x* directions), leading to anisotropic spin conductivity, while the overall conductivity of the system remains constant. To further validate our analysis, we performed first principles quantum transport simulations based on NEGF-DFT. A two-probe device was constructed (Fig. 4a), comprising a central scattering region based on bilayer $V_2S_2O$ with semi-infinite Au electrodes in direct contact with the $V_2S_2O$ surface. Electron transport is investigated along the $x$ direction. A gate voltage ($E_\perp$) perpendicular to the bilayer $V_2S_2O$ is applied in the transport calculations, with magnitudes of 0 and ±0.8 V.

Figure 4b displays the spin-resolved transmission spectra of bilayer $V_2S_2O$ under gated and gate-free conditions. The Fermi level ($E_F$) is set to 0 eV as the energy reference, which corresponds to the Fermi level of the semi-infinite Au electrodes. Due to the formation of a Schottky barrier at the Au/$V_2S_2O$ interface, this reference energy point lies within the band gap of bilayer $V_2S_2O$, as evidenced by the vanishing transmission coefficient at $E_F$ = 0 eV. It is evident that the device exhibits a pronounced spin current above the Fermi level regardless of the gate voltage, whereas the spin current becomes negligible for $E_F$ < 0. This transport behavior is closely associated with the competition in the VBM position between the $p_z$ and $p_{xy}$ orbitals, which arises from interlayer interactions in the bilayer system. Since Au electrodes are capped exclusively on the top layer of $V_2S_2O$, the top layer is expected to dominate the total spin current contribution in principle. In the absence of interlayer interactions, the spin polarization of the transmission coefficient for bilayer $V_2S_2O$ should be comparable to that of the monolayer counterpart. However, the calculated results reveal a remarkably lower spin polarization in the bilayer system than in the monolayer (Fig. S5): for instance, at $E_F$ = 1 eV, the spin polarization ratio of the transmission coefficient is only 63% for the bilayer, in stark contrast to the nearly 100% achieved for the monolayer. This significant discrepancy demonstrates that interlayer interactions drastically modulate the spin transport properties of bilayer $V_2S_2O$.

On the other hand, the gate voltage is found to effectively tune the charge-to-spin current conversion efficiency of the bilayer device: an applied gate voltage of +0.8 V lifts the spin polarization ratio of the transmission coefficient to 72% at $E_F$ = 1 eV. Notably, the bilayer system exhibits a distinct bias effect in response to gate voltage modulation: a gate voltage of -0.8 V results in a spin polarization ratio of 62% at $E_F$ = 1 eV, merely a 1% reduction relative to the gate-free case. This directional dependence of the gate voltage effect stems from the asymmetric transport contributions of the top and bottom layers of $V_2S_2O$. As illustrated in Fig. 3, a positive out-of-plane electric field significantly lowers the CBM of the bottom layer, thereby enhancing its contribution to the overall low-energy device transport. In contrast, a negative gate voltage suppresses the transport contribution of the bottom layer; yet, owing to the intrinsically minor role of the bottom layer in the total transport, this suppression only leads to a slight reduction in the spin polarization ratio of the total transmission coefficient.

**CONCLUSIONS**

In summary, we systematically studied the profound impacts of interlayer interactions on the electronic, magnetic and quantum transport properties of bilayer altermagnetic $V_2S_2O$ via first-principles simulations. Interlayer coupling selectively modulating p-orbital-dominated top valence bands to induce a tiny energy competition for VBM between Γ-point $p_z$ and X/Y-point $p_{xy}$ orbitals, and imposing compressive strain/hole doping dual requirements for a robust piezomagnetic effect—distinct from the strain-insensitive monolayer. Out-of-plane electric fields drive a large Stark shift, weaken interlayer coupling, and can convert the normal AFM bilayer to a layer-resolved altermagnetic phase. Quantum transport simulations show interlayer interactions significantly reduce transmission spin polarization, with vertical gate voltage exhibiting a remarkable yet seemingly inevitable asymmetric modulation effect. Our findings elucidate interlayer coupling effects in 2D altermagnets, providing critical insights for designing high-performance multilayer altermagnetic spintronic devices with external field tunability.

**METHODS**
First-principles calculations were performed using the Vienna *ab initio* simulation package (VASP) [36-37]. The exchange-correlation functional adopts the Perdew-Burke-Ernzerhof (PBE) form within generalized gradient approximation (GGA). An 8×8×1 Γ-centered *k* grid, a plane-wave energy cutoff of 600 eV, and the Gaussian smearing method with a width of 0.05 eV were adopted. A vacuum layer of around 25 Å was introduced to reduce the periodic interactions, while the van der Waals (vdW) effect was corrected using the DFT-D3 method [38]. A effective Hubbard method [39] with U value of 4 eV was used to reduce the seif-interaction errors of localized (strongly correlated) 3d orbital of V. Transport calculations were performed by using the quantum transport software NANODCAL [40] within the NEGF formalism [41-42]. The double zeta polarized (DZP) atomic orbital basis is used to expand all the physical quantities with an energy cutoff of 80 Hartree; the exchange-correlation functional was adopted at the PBE level as well; atomic cores were represented by standard norm-conserving nonlocal pseudopotentials; a Fermi smearing of 0.05 eV and 10 *k* points were used for the transport direction.

**ASSOCIATED CONTENT**
Data Availability Statement

The data that support the findings of this article can be acquired by contacting the corresponding author.

**Author Contributions**
The manuscript was written through contributions of all authors. All authors have given approval to the final version of the manuscript. Notes The authors declare no competing financial interest.

**ACKNOWLEDGMENTS**
Work at HDU was supported by Zhejiang Provincial Natural Science Foundation (LQN25A040011), National Natural Science Foundation of China (12504264, 12104118), and the Foundation of Hangzhou Dianzi University (KYS075624288). We gratefully acknowledge HZWTECH for providing computation facilities.

**REFERENCES**

1. Jungwirth, T.; Marti, X.; Wadley, P.; Wunderlich, J., Antiferromagnetic spintronics. *Nature Nanotechnology* **2016,** *11* (3), 231-241.
2. Žutić, I.; Fabian, J.; Das Sarma, S., Spintronics: Fundamentals and applications. *Reviews of Modern Physics* **2004,** *76* (2), 323-410.
3. Šmejkal, L.; Sinova, J.; Jungwirth, T., Emerging research landscape of altermagnetism. *Physical Review X* **2022,** *12* (4), 040501.
4. Baltz, V.; Manchon, A.; Tsoi, M.; Moriyama, T.; Ono, T.; Tserkovnyak, Y., Antiferromagnetic spintronics. *Reviews of Modern Physics* **2018,** *90* (1), 015005.
5. Šmejkal, L.; Sinova, J.; Jungwirth, T., Beyond Conventional Ferromagnetism and Antiferromagnetism: A Phase with Nonrelativistic Spin and Crystal Rotation Symmetry. *Physical Review X* **2022,** *12* (3), 031042.
6. Zhang, X.; Liu, Q.; Luo, J.-W.; Freeman, A. J.; Zunger, A., Hidden spin polarization in inversion-symmetric bulk crystals. *Nature Physics* **2014,** *10* (5), 387-393.
7. Liu, P.; Li, J.; Han, J.; Wan, X.; Liu, Q., Spin-Group Symmetry in Magnetic Materials with Negligible Spin-Orbit Coupling. *Physical Review X* **2022,** *12* (2), 021016.
8. Reichlova, H.; Lopes Seeger, R.; González-Hernández, R.; Kounta, I.; Schlitz, R.; Kriegner, D.; Ritzinger, P.; Lammel, M.; Leiviskä, M.; Birk Hellenes, A., Observation of a spontaneous anomalous Hall response in the Mn5Si3 d-wave altermagnet candidate. *Nature Communications* **2024,** *15* (1), 4961.
9. Chen, H.; Liu, L.; Zhou, X.; Meng, Z.; Wang, X.; Duan, Z.; Zhao, G.; Yan, H.; Qin, P.; Liu, Z., Emerging antiferromagnets for spintronics. *Advanced Materials* **2024,** *36* (14), 2310379.
10. Ma, H.-Y.; Hu, M.; Li, N.; Liu, J.; Yao, W.; Jia, J.-F.; Liu, J., Multifunctional antiferromagnetic materials with giant piezomagnetism and noncollinear spin current. *Nature communications* **2021,** *12* (1), 2846.
11. Samanta, K.; Shao, D.-F.; Tsymbal, E. Y., Spin filtering with insulating altermagnets. *Nano Letters* **2025,** *25* (8), 3150-3156.
12. González-Hernández, R.; Šmejkal, L.; Výborný, K.; Yahagi, Y.; Sinova, J.; Jungwirth, T.; Železný, J., Efficient Electrical Spin Splitter Based on Nonrelativistic Collinear Antiferromagnetism. *Physical Review Letters* **2021,** *126* (12), 127701.
13. Liu, C.; Li, X.; Li, X.; Yang, J., Realizing Abundant Two-Dimensional Altermagnets with Anisotropic Spin Current Via Spatial Inversion Symmetry Breaking. *Nano Letters* **2025,** *25* (23), 9197-9203.
14. Reimers, S.; Odenbreit, L.; Šmejkal, L.; Strocov, V. N.; Constantinou, P.; Hellenes, A. B.; Jaeschke Ubiergo, R.; Campos, W. H.; Bharadwaj, V. K.; Chakraborty, A.; Denneulin, T.; Shi, W.; Dunin-Borkowski, R. E.; Das, S.; Kläui, M.; Sinova, J.; Jourdan, M., Direct observation of altermagnetic band splitting in CrSb thin films. *Nature Communications* **2024,** *15* (1), 2116.
15. Fedchenko, O.; Minár, J.; Akashdeep, A.; D'Souza, S. W.; Vasilyev, D.; Tkach, O.; Odenbreit, L.; Nguyen, Q.; Kutnyakhov, D.; Wind, N.; Wenthaus, L.; Scholz, M.; Rossnagel, K.; Hoesch, M.; Aeschlimann, M.; Stadtmüller, B.; Kläui, M.; Schönhense, G.; Jungwirth, T.; Hellenes, A. B.; Jakob, G.; Šmejkal, L.; Sinova, J.; Elmers, H.-J., Observation of time-reversal symmetry breaking in the band structure of altermagnetic RuO2. *Science Advances* **2024,** *10* (5), eadj4883.
16. Jiang, B.; Hu, M.; Bai, J.; Song, Z.; Mu, C.; Qu, G.; Li, W.; Zhu, W.; Pi, H.; Wei, Z.; Sun, Y.-J.; Huang, Y.; Zheng, X.; Peng, Y.; He, L.; Li, S.; Luo, J.; Li, Z.; Chen, G.; Li, H.; Weng, H.; Qian, T., A metallic room-temperature d-wave altermagnet. *Nature Physics* **2025,** *21* (5), 754-759.
17. Sun, Y.; Huang, Y.; Cheng, J.; Zhang, S.; Li, Z.; Luo, H.; Ma, X.; Yang, W.; Yang, J.; Chen, D.; Sun,

17. K.; Gutmann, M.; Capelli, S. C.; Shen, F.; Hao, J.; He, L.; Chen, G.; Li, S., Antiferromagnetic structure of KV2Se2O: A neutron diffraction study. *Physical Review B* **2025,** *112* (18), 184416.

18. Liu, B.; Fu, P.-H.; Sun, Y.-X.; Zhang, X.-L.; Zhu, S.-C.; Yu, X.-L.; Wu, H.; Shao, Y.-Z., Intrinsic Spin Filter Effect in a $ d $-wave altermagnet KV2Se2O with Open Fermi Surface. *arXiv preprint arXiv:2602.21460* **2026**.

19. Scheike, T.; Xiang, Q.; Wen, Z.; Sukegawa, H.; Ohkubo, T.; Hono, K.; Mitani, S., Exceeding 400% tunnel magnetoresistance at room temperature in epitaxial Fe/MgO/Fe(001) spin-valve-type magnetic tunnel junctions. *Applied Physics Letters* **2021,** *118* (4).

20. Yuasa, S.; Nagahama, T.; Fukushima, A.; Suzuki, Y.; Ando, K., Giant room-temperature magnetoresistance in single-crystal Fe/MgO/Fe magnetic tunnel junctions. *Nature Materials* **2004,** *3* (12), 868-871.

21. Wang, X.; Lyu, K.-Y.; Guo, Y.; Li, Y.-X., Quantized crossed Andreev reflection in altermagnet/altermagnet topological superconductor/altermagnet heterojunction. *Physical Review B* **2026,** *113* (1), 014505.

22. Wang, D.; Ghosh, A. K.; Tao, Y.; Ma, F.; Song, C., Emerging of Anomalous Higher-Order Topological Phases in Altermagnet/Topological Insulator Heterostructure by Floquet Engineering. *Advanced Science* **2026**, e22203.

23. Pan, B.; Zhou, P.; Lyu, P.; Xiao, H.; Yang, X.; Sun, L., General Stacking Theory for Altermagnetism in Bilayer Systems. *Physical Review Letters* **2024,** *133* (16), 166701.

24. Sun, W.; Wang, W.; Yang, C.; Hu, R.; Yan, S.; Huang, S.; Cheng, Z., Altermagnetism Induced by Sliding Ferroelectricity via Lattice Symmetry-Mediated Magnetoelectric Coupling. *Nano Letters* **2024,** *24* (36), 11179-11186.

25. Zhang, R.-W.; Cui, C.; Li, R.; Duan, J.; Li, L.; Yu, Z.-M.; Yao, Y., Predictable gate-field control of spin in altermagnets with spin-layer coupling. *Physical Review Letters* **2024,** *133* (5), 056401.

26. Li, J.-Y.; Fan, A.-D.; Wang, Y.-K.; Zhang, Y.; Li, S., Strain-induced valley polarization, topological states, and piezomagnetism in two-dimensional altermagnetic V2Te2O, V2STeO, V2SSeO, and V2S2O. *Applied Physics Letters* **2024,** *125* (22).

27. Wang, D.; Wang, H.; Liu, L.; Zhang, J.; Zhang, H., Electric-field-induced switchable two-dimensional altermagnets. *Nano Letters* **2024,** *25* (1), 498-503.

28. Aoyama, T.; Ohgushi, K., Piezomagnetic properties in altermagnetic MnTe. *Physical Review Materials* **2024,** *8* (4), L041402.

29. Zhu, Y.; Chen, T.; Li, Y.; Qiao, L.; Ma, X.; Liu, C.; Hu, T.; Gao, H.; Ren, W., Multipiezo effect in altermagnetic V2SeTeO monolayer. *Nano Letters* **2023,** *24* (1), 472-478.

30. Gu, M.; Liu, Y.; Zhu, H.; Yananose, K.; Chen, X.; Hu, Y.; Stroppa, A.; Liu, Q., Ferroelectric Switchable Altermagnetism. *Physical Review Letters* **2025,** *134* (10), 106802.

31. Ga, Y.; Zhang, F.; Wang, L.; Jiang, J.; Chang, K.; Yang, H., Interlayer exchange coupling driven magnetic phase transition in a two-dimensional lattice. *Physical Review B* **2025,** *112* (2), L020407.

32. Peng, R.; Yang, J.; Hu, L.; Ong, W.-L.; Ho, P.; Lau, C. S.; Liu, J.; Ang, Y. S., All-electrical layer-spintronics in altermagnetic bilayers. *Materials Horizons* **2025,** *12* (7), 2197-2207.

33. Qi, Y.; Zhao, J.; Zeng, H., Spin-layer coupling in two-dimensional altermagnetic bilayers with tunable spin and valley splitting properties. *Physical Review B* **2024,** *110* (1), 014442.

34. Bandyopadhyay, S.; Atulasimha, J.; Barman, A., Magnetic straintronics: Manipulating the magnetization of magnetostrictive nanomagnets with strain for energy-efficient applications. *Applied Physics Reviews* **2021,** *8* (4).

35. Yuan, J.-H.; Yang, X.-K.; Wei, B.; Chen, Y.-B.; Cui, H.-Q.; Liu, J.-H.; Dou, S.-Q.; Song, M.-X.; Fei, L., Strain-Mediated Magnetization Switching Behavior in a Bicomponent Nanomagnet. *Physical Review Applied* **2023,** *19* (1), 014003.
36. Kresse, G.; Furthmüller, J., Efficiency of ab-initio total energy calculations for metals and semiconductors using a plane-wave basis set. *Computational Materials Science* **1996,** *6* (1), 15-50.
37. Kresse, G.; Furthmüller, J., Efficient iterative schemes for ab initio total-energy calculations using a plane-wave basis set. *Physical Review B* **1996,** *54* (16), 11169-11186.
38. Grimme, S.; Ehrlich, S.; Goerigk, L., Effect of the damping function in dispersion corrected density functional theory. *Journal of computational chemistry* **2011,** *32* (7), 1456-1465.
39. Dudarev, S. L.; Botton, G. A.; Savrasov, S. Y.; Humphreys, C. J.; Sutton, A. P., Electron-energy-loss spectra and the structural stability of nickel oxide: An LSDA+U study. *Physical Review B* **1998,** *57* (3), 1505-1509.
40. Xie, Y.; Chen, M.; Wu, Z.; Hu, Y.; Wang, Y.; Wang, J.; Guo, H., Two-Dimensional Photogalvanic Spin-Battery. *Physical Review Applied* **2018,** *10* (3).
41. Henrickson, L. E. J. J. o. a. p., Nonequilibrium photocurrent modeling in resonant tunneling photodetectors. **2002,** *91* (10), 6273-6281.
42. Xie, Y.; Zhang, L.; Zhu, Y.; Liu, L.; Guo, H. J. N., Photogalvanic effect in monolayer black phosphorus. **2015,** *26* (45), 455202.